\documentclass[11pt,a4paper]{article}

\usepackage{jheppub}
\usepackage{epstopdf}

\usepackage{mathrsfs}
\usepackage{psfrag}
\usepackage{color}
\usepackage{hyperref}

\usepackage{slashed}
\usepackage{simplewick}
\usepackage{cancel}
\usepackage[utf8]{inputenc}

\usepackage[table]{xcolor}

\usepackage{amsmath,bbm,array,amsfonts,graphicx,wrapfig,arydshln,lscape,float,multirow,longtable,rotating,makecell}
\usepackage{url}

\usepackage{caption}
\usepackage{subcaption}

\usepackage[backgroundcolor=black!5, linecolor=green!60]{todonotes}

\hypersetup{pdftitle={},pdfcreator={},linkcolor=[rgb]{0.15,0.35,0.75},colorlinks=true,citecolor=[rgb]{0.675,0,0.2},urlcolor=[rgb]{0.15,0.35,0.65}}
\let\olditemize\itemize\renewcommand{\itemize}{\vspace{-2pt}\olditemize\setlength{\itemsep}{1pt}\setlength{\parskip}{0pt}\setlength{\parsep}{-0pt}}
\let\oldenumerate\enumerate\renewcommand{\enumerate}{\vspace{-4pt}\oldenumerate\setlength{\itemsep}{1pt}\setlength{\parskip}{0pt}\setlength{\parsep}{0pt}}

\renewcommand{\thefootnote}{\fnsymbol{footnote}}
\newpage
\csname @addtoreset\endcsname{equation}{section}

\makeatletter
\def\timenow{\@tempcnta\time
  \@tempcntb\@tempcnta
  \divide\@tempcntb60
  \ifnum10>\@tempcntb0\fi\number\@tempcntb
  \multiply\@tempcntb60
  \advance\@tempcnta-\@tempcntb
  :\ifnum10>\@tempcnta0\fi\number\@tempcnta}
\makeatother

\allowdisplaybreaks[3]

\makeatletter
\renewcommand\@fpheader{} 
\renewcommand\@journal{}
\makeatother

\title{
Conformal anomaly of pseudo-scalar operators
}

\preprint{LAPTH-008/20, MPP-2020-32}

\author{D.\ Chicherin$^a$, E.\ Sokatchev$^{b}$}

\affiliation{
$^a$ Max-Planck-Institut f{\"u}r Physik, Werner-Heisenberg-Institut, 80805 M{\"u}nchen, Germany\\
$^b$ LAPTh, Universit\'e Savoie Mont Blanc, CNRS, B.P. 110, F-74941 Annecy-le-Vieux, France}

\emailAdd{chicheri@mpp.mpg.de}
\emailAdd{emeri.sokatchev@cern.ch}

\abstract{
We study the conformal properties of pseudo-scalar (parity-odd) operators in four-dimensional field theory. Such operators include the topological term in a gauge theory and the Yukawa coupling. We show that a single operator of this type cannot be a conformal primary on its own. This is only possible for certain mixtures of pseudo-scalar operators like the chiral Lagrangian of ${\cal N}=4$ super-Yang-Mills  theory. We examine the detailed mechanism of conformal symmetry breakdown and derive an anomalous conformal Ward identity. We explore a possible link between this conformal anomaly and the axial anomaly. }

\begin{document}

\maketitle

\renewcommand{\thefootnote}{\fnsymbol{footnote}}
\pagestyle{empty}
\setcounter{page}{0}


\newcommand{\norm}[1]{{\protect\normalsize{#1}}}
\newcommand{\p}[1]{(\ref{#1})}
\newcommand{\half}{{\ts \frac{1}{2}}}
\newcommand \vev [1] {\langle{#1}\rangle}
\newcommand \ket [1] {|{#1}\rangle}
\newcommand \bra [1] {\langle {#1}|}

\newcommand{\cI}{{\cal I}}
\newcommand{\cM}{{\cal M}} 
\newcommand{\cR}{{\cal R}} 
\newcommand{\cS}{{\cal S}} 
\newcommand{\cK}{{\cal K}}
\newcommand{\cL}{{\cal L}} 
\newcommand{\cF}{{\cal F}}
\newcommand{\cN}{{\cal N}}
\newcommand{\cA}{{\cal A}}
\newcommand{\cB}{{\cal B}}
\newcommand{\cG}{{\cal G}}
\newcommand{\cO}{{\cal O}}
\newcommand{\cY}{{\cal Y}}
\newcommand{\cX}{{\cal X}}
\newcommand{\cT}{{\cal T}}
\newcommand{\cD}{{\cal D}}
\newcommand{\cW}{{\cal W}}
\newcommand{\cP}{{\cal P}}
\newcommand{\mK}{{\mathbb K}}
\newcommand{\nt}{\notag\\} 
\newcommand{\pa}{\partial}
\newcommand{\ep}{\epsilon}
\newcommand{\om}{\omega}
\newcommand{\bep}{\bar\epsilon}
\newcommand{\vep}{\varepsilon}
\renewcommand{\a}{\alpha}
\renewcommand{\b}{\beta}
\renewcommand{\d}{\delta}
\newcommand{\g}{\gamma}
\newcommand{\s}{\sigma}
\newcommand{\la}{\lambda}
\newcommand{\tl}{\tilde\lambda}
\newcommand{\tm}{\tilde\mu}
\newcommand{\tk}{\tilde k}
\newcommand{\da}{{\dot\alpha}}
\newcommand{\db}{{\dot\beta}}
\newcommand{\dg}{{\dot\gamma}}
\newcommand{\dd}{{\dot\delta}}
\newcommand{\q}{\theta}
\newcommand{\bq}{\bar\theta}
\newcommand{\Q}{\Theta}
\newcommand{\bQ}{\bar Q}
\newcommand{\bz}{\bar z}
\newcommand{\tx}{\tilde{x}}
\newcommand{\tr}{\mbox{tr}}
\newcommand{\+}{{\dt+}}
\renewcommand{\-}{{\dt-}}
\newcommand{\ti}{{\textup{i}}}


\setcounter{page}{1}\setcounter{footnote}{0}

\pagestyle{plain}
\renewcommand{\thefootnote}{\arabic{footnote}}



\setcounter{page}{1}\setcounter{footnote}{0}


\section{Introduction}\label{s1}

Conformal symmetry is a powerful constraint on the dynamics of quantum field theories. The theories with exact conformal symmetry have a vanishing beta function, e.g., $\cN=4$ supersymmetric Yang-Mills theory (SYM). Conformal symmetry also has numerous applications in QCD,   at leading order where the symmetry is unbroken, and beyond (see \cite{Braun:2003rp} for a review).  

 A key concept in conformal symmetry is that of a conformal primary operator $O(x)$ satisfying the defining condition $\mathbb{K}_\mu O(0)=0$, where $\mathbb{K}_\mu$ is the generator of special conformal transformations.  The  correlation functions of such operators are severely constrained by conformal invariance and by the requirement of crossing symmetry (the so-called `bootstrap', see \cite{Poland:2018epd} for a review). In a four-dimensional field theory the operators are classified according to their parity properties into proper (parity-even) and pseudo (parity-odd) scalars, vectors, tensors, etc.  
In this note we study the conformal properties of pseudo-scalar  operators. We show that a single operator of this type cannot be a conformal primary on its own, because of a specific conformal anomaly. This is only possible for certain anomaly-free mixtures of such operators, such as the Lagrangian of the $\cN=4$ SYM theory. We also make contact between this new conformal anomaly and the well-known axial anomaly in non-supersymmetric theories.  

The best known example of a pseudo-scalar  operator is the topological term $\Q = \frac{1}{2}\ep_{\mu\nu\la\rho} \tr(F^{\mu\nu} F^{\la\rho}) \equiv F\tilde F$ built from the curvature $F_{\mu\nu}$ of a gauge theory, such as massless QCD/QED. Even though such theories have a non-vanishing beta function and hence broken conformal invariance, this effect will only show up at two loops (order $\sim g^4$ in the coupling). At one loop (order $\sim g^2$) the operator $\Q$ may need renormalization (in the non-Abelian case) and may acquire an anomalous dimension. Yet, this is not in conflict with the renormalized operator being a conformal primary. What really makes this impossible is the parity-odd nature of the operator. Here is a simple argument explaining the point.

If an operator is a conformal primary (in a theory with a vanishing beta function), its correlation functions with other conformal operators should be conformally covariant. Examples of well behaved conformal operators are the conserved currents, e.g., electromagnetic current, energy-momentum tensor, etc. They are protected form renormalization and keep their canonical dimension. So, let us consider the three-point correlator
\begin{align}\label{1.1}
\vev{V_\mu(x_1) V_\nu(x_2) \Q(x_3) } =  \ep_{\mu\nu\la\rho} x^\la_{12} x^\rho_{13} \, F(x^2_{ij}) \,,
\end{align}
where $V_\mu(x)= \tr(\bar\Psi\g_\mu  \Psi)$ is an electromagnetic current and $x_{ij} = x_i - x_j$. The expression on the right-hand side reflects Poincar\'e invariance as well as the parity property of the correlator. Indeed, $\Q$ being parity-odd and $V$ being a proper vector (parity-even), the whole objects must be a pseudo-tensor of rank two, hence the presence of the Levi-Civita tensor. The further properties of the correlator, such as conservation at points 1 and 2 and its scaling dimension, have not yet been implemented in \p{1.1}. For the purpose of our argument, we are only interested in its conformal covariance. If this object were conformal, we could choose the conformal frame $x_2=0$ and $x_3=\infty$ (after compensating the conformal weight at point 3 by an appropriate factor). Then we would be left with the single vector $x^\mu_1$, so the expression \p{1.1} would vanish. Furthermore, if a conformal correlation function vanishes in a particular conformal frame, it vanishes in any frame. We conclude that the only way for the correlator \p{1.1} to be conformal is  to vanish identically.

Now, it is easy to see that in a theory with a gauge field and fermion matter, such as massless QED or QCD, the one-loop (order $\sim g^2$) three-point function \p{1.1} does not vanish (see Section~\ref{secBL3pC} for a detailed calculation). This implies that the operator $\Q$ is not  a conformal primary. We could try to repair this `defect' by allowing $\Q$ to mix with some other operators with the same properties (pseudo-scalar of dimension 4). In QED/QCD there is such an operator, the divergence of the axial current $A_\mu = \tr(\bar\Psi\g_\mu \g_5 \Psi)$. The two operators do indeed mix but only starting at two-loop level \cite{Larin:1993tq}. The situation improves in theories with fermion and scalar matter, such as $\cN=4$ SYM. The latter theory has a vanishing beta function, so it makes sense to consider conformal primary operators at any perturbative level. In this case there exists another operator of the same type, the pseudo-scalar Yukawa coupling $\cY=\tr(\phi\bar\Psi\g_5 \Psi)$ where $\phi$ is the scalar matter field. As we show in Section~\ref{secSusy}, a particular combination $\Q+g\cY$ is indeed a well-defined conformal primary. 

We can ask the question: What is the reason why the operator $\Q$ cannot be a conformal primary? As noted earlier, at the lowest perturbative level this cannot be the breakdown of conformal invariance due to the beta function, nor an ultraviolet renormalization artifact. It turns out that the operator has a hidden singularity when inserted into a fermion propagator. The singularity becomes visible only if we make a conformal transformation of the Feynman diagram, and it has the effect of producing an {\it anomalous conformal Ward identity}. This somewhat subtle mechanism is analyzed in detail in Section~\ref{secAnom}. Such an anomaly is not limited to the three-point function \p{1.1}, it is present in the correlator of $\Q$ with any number of vector currents. In contrast, the anomaly does not occur for the parity-even scalar operator $\cL= -\frac{1}{2}\tr(F^{\mu\nu} F_{\mu\nu})$, which is the gauge field Lagrangian. The distinction between the scalar and pseudo-scalar operators is best seen using two-component spinor (chiral) notation (see Appendix~\ref{AppConv}).\footnote{Whenever a regularization is  needed, we use a version of the dimensional reduction scheme \cite{Siegel:1979wq,Bern:1995db} in which the chiral notation is justified. } In it the chiral Lagrangian $L=-\frac{1}{2}\tr(F^{\a\b} F_{\a\b})$ is complex, and its complex conjugate $\bar L = -\frac{1}{2}\tr(\tilde F^{\da\db} \tilde F_{\da\db})$ is the anti-chiral Lagrangian. The real and imaginary parts 
\begin{align}\label{}
\cL= \frac1{2}(L+\bar L) \,, \qquad \Q= \frac1{i}(L-\bar L)
\end{align}
are the standard Yang-Mills Lagrangian and the topological term, respectively. The conformal anomaly occurs when inserting the chiral Lagrangian (or its anti-chiral conjugate) in a fermion line but it cancels if the real part is inserted (see Section \ref{secAnom}). In a theory with scalar matter, such as $\cN=4$ SYM, the insertion of the chiral Yukawa coupling $Y=\tr(\phi \psi^\a \psi_\a)$ generates a similar anomaly, so that the chiral combination $L+gY$ is anomaly free. This combination (completed with the real $\phi^4$ term) happens to be the chiral on-shell Lagrangian of the $\cN=4$ SYM theory, a component of the energy-momentum tensor supermultiplet. 

Finally, in Section \ref{secChirAnom} we comment on a possible link between our conformal anomaly and the well-known axial anomaly. The axial current $A_\mu$ is not conserved at loop level, therefore it acquires an anomalous dimension. The divergence $\pa^\mu A_\mu$ of a vector of non-canonical dimension is a conformal descendant, not a conformal primary. Since this divergence is related to the topological term $\Q$ by the Adler-Bardeen theorem, the latter cannot be a conformal primary either.

\section{Born-level  correlator of two vector currents and the topological term}  \label{s2}

In this Section we calculate a three-point correlation function of gauge-invariant composite operators in the Born approximation (the lowest perturbative order) and find that it is not conformal despite of the classical conformal invariance of the underlying theory and of the composite operators. Our argument is valid for any 4D massless gauge theory involving fermions (e.g. massless QED, QCD, or super-Yang-Mills theory); non-abelian effects do not appear at this perturbative level. Only the following part of the Lagrangian is relevant in our Feynman graph calculations,
\begin{align} 
L_{\rm QCD} =  - \frac{1}{2} \tr \, F_{\mu \nu} F^{\mu \nu} + \frac{i}{2}\, \bar\Psi  \gamma^\mu \overset{\leftrightarrow}{\cal D}_{\mu} \Psi \,,
\label{Lagr}
\end{align}
where ${\cal D}_\mu = \pa_\mu - i g \cA_\mu$ is the covariant derivative with $SU(N_c)$ gauge connection $\cA_{\mu} = \cA^a_{\mu} T_a$ and $F_{\mu \nu} = \frac{i}{g}[{\cal D}_\mu , {\cal D}_\nu]$ is the field-strength tensor. The color generators are normalized as $\tr(T_a T_b) = \frac{1}{2}\delta_{ab}$. The Dirac spinor $\Psi_{i}$ and its conjugate $\bar\Psi^i$ are in the (anti)-fundamental representation of $SU(N_c)$, i.e. $\bar\Psi^i(T_a)_{i}{}^{j} \Psi_j$. The Lagrangian is invariant under conformal transformations classically; the UV renormalization of the fields and of the coupling constant do not appear at this perturbative level.

In view of the global $U(1)\times U(1)$ invariance of the Lagrangian \p{Lagr} 
the electromagnetic vector $V_{\mu} =\tr \,  \bar\Psi\gamma_\mu \Psi$
and axial vector $A_{\mu} =\tr \,  \bar\Psi\gamma_\mu\gamma_5 \Psi$ currents are classically conserved. The vector current $V_{\mu}$ does not require infinite UV renormalizations and it is conserved at the quantum level as well. The conservation of the axial current at the quantum level is spoiled by the Adler-Bardeen anomaly which is one-loop exact.

Let us also introduce the dual field strength tensor $\tilde F_{\mu\nu} = \frac{1}{2}\epsilon_{\mu\nu\rho\la} F^{\rho\la}$ and consider the following pseudo-scalar gauge-invariant operator  
\begin{align}
\Theta = \tr\, F_{\mu\nu} \tilde F^{\mu\nu} \,, \label{Theta}
\end{align}
which is the well-know topological term, the divergence of a the gauge non-invariant Chern-Simons term.

\begin{figure}
\begin{center}
\begin{tabular}{ccc}
\begin{tabular}{c}
\includegraphics[width=4.8cm]{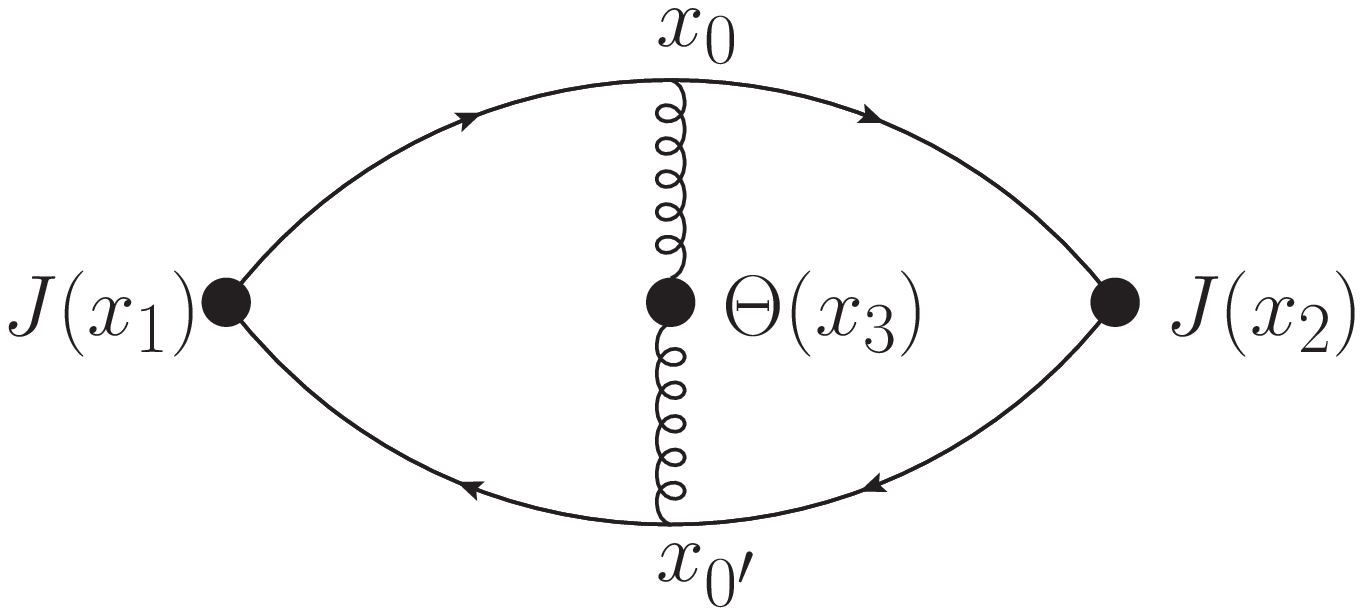}
\end{tabular} & 
\begin{tabular}{c}
\includegraphics[width=4.2cm]{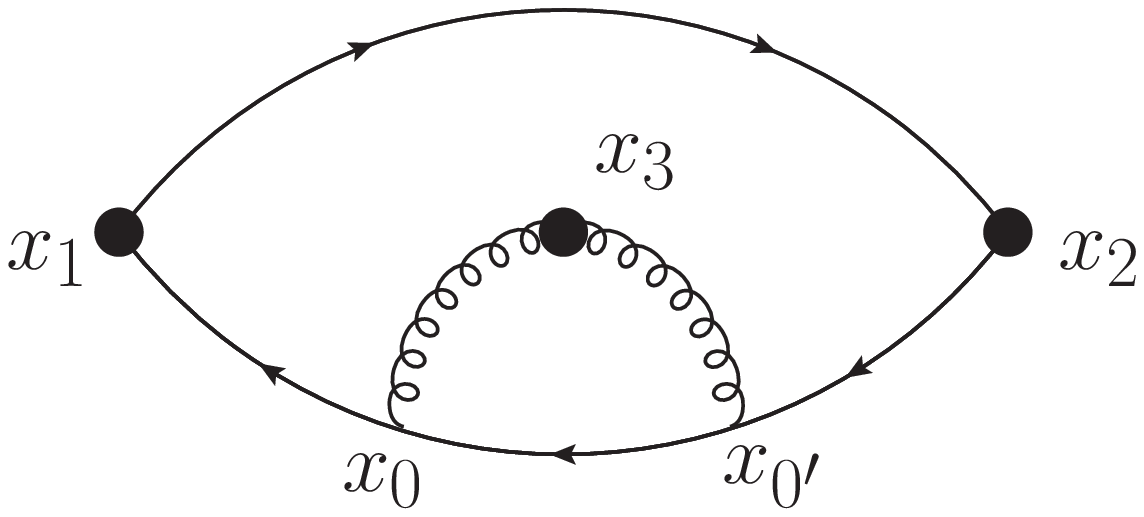}
\end{tabular} &
\begin{tabular}{c}
\includegraphics[width=4.2cm]{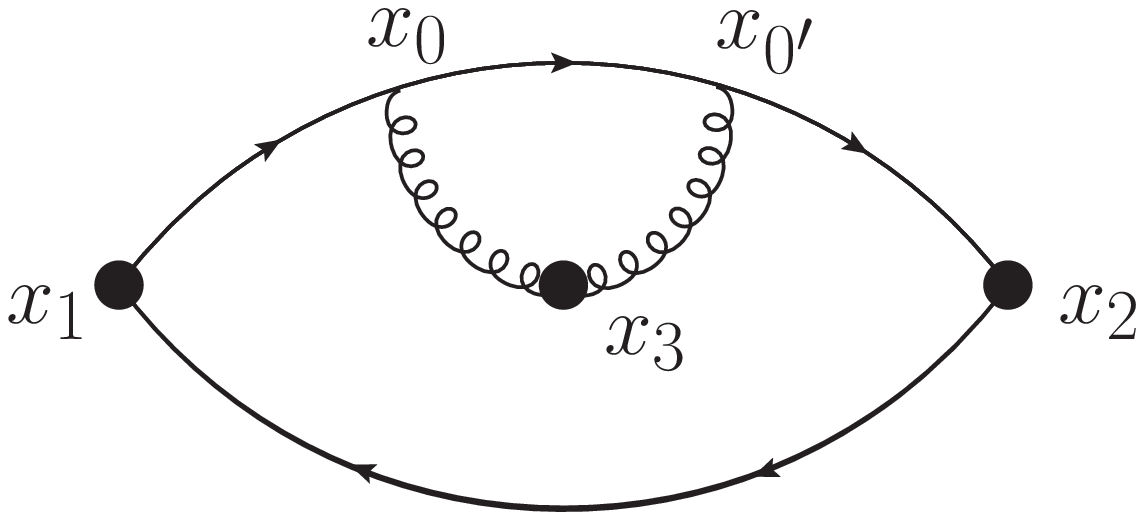}\\
{}
\end{tabular}
\end{tabular}
\end{center}
\caption{Feynman diagrams for the Born-level correlator $\vev{J_\mu(x_1)J_\nu(x_2)\Theta(x_3)}_{g^2}$.}\label{3pt}
\end{figure}

The currents $V_\mu$ and the pseudo-scalar $\Theta$ are classically conformally covariant operators. Let us consider their three-point correlation function in the lowest perturbative approximation
\begin{align}
G^{(\rm odd)}_{\mu\nu}(x_1,x_2,x_3) =\vev{V_\mu(x_1)\, V_\nu(x_2) \, \Theta(x_3) }_{\rm Born} \,. \label{corrJJTh}
\end{align}
The relevant Feynman diagrams are shown in Fig.~\ref{3pt}. In this case the lowest perturbative order is ${\cal O}(g^2)$. Despite of the fact that the diagrams involve space-time integration vertices ($x_0$ and $x_{0'}$) their sum turns out to be a rational function.\footnote{Using the amplitude terminology we could say that the correlator is zero at the tree-level ${\cal O}(g^0)$, and its perturbative expansion starts at one-loop ${\cal O}(g^2)$.}

\subsection{Chiral and anti-chiral Lagrangian insertions}

In the following we prefer to use two-component Lorentz spinor index notations, see App.~\ref{AppConv}. We decompose the Dirac fermion in a pair of Weyl fermions $\chi$ and $\psi$,
\begin{align}
\Psi = (\chi_{\a} \, ,\, \bar\psi^{\da}) \;\;\;,\;\;\;\;\;\;
\bar\Psi = (\psi^{\a} \, , \, \bar\chi_{\da}) \,.
\end{align}
The vector and axial vector currents are independent linear combinations of two Majorana currents,
\begin{align}
& V_{\mu} = \psi^{\a} \sigma^\mu_{\a\da} \bar\psi^{\da} - \chi^{\a} \sigma^\mu_{\a\da} \bar\chi^{\da}= \bar\Psi\g_\mu\Psi\,,  \label{2.5}\\
&A_{\mu} = \psi^{\a} \sigma^\mu_{\a\da} \bar\psi^{\da} + \chi^{\a} \sigma^\mu_{\a\da} \bar\chi^{\da} = i\bar\Psi\g_\mu\g_5\Psi \,.  \label{2.6}\
\end{align}
With a single Majorana spinor $\Psi^*=\Psi$ or equivalently $\chi=\psi$, only the axial vector combination \p{2.6} is possible. For the purpose of the Feynman graph calculations in this section we use just the real current $J_{\a\da} \equiv \sigma^{\mu}_{\a\da} J_\mu = \psi_{\a} \bar\psi_{\da}$ made of a single Majorana spinor $(\psi,\bar\psi)$; adding or subtracting the contribution of the other Majorana spinor $(\chi,\bar\chi)$ is straightforward. 

We also split the real field strength tensor $F_{\mu\nu}$ into its chiral (or self-dual) $F_{\a\b} = F_{\b\a}$ and anti-chiral (or anti-self-dual) $\tilde F_{\da\db} = \tilde F_{\db\da}$ components
\begin{align}
F_{\mu\nu}\, \sigma^{\mu}_{\a\da} \sigma^{\nu}_{\b\db} & = \epsilon_{\a\b} \tilde F_{\da\db} + \epsilon_{\da\db} F_{\a\b}  \,, \\
\tilde F_{\mu\nu}\, \sigma^{\mu}_{\a\da} \sigma^{\nu}_{\b\db} & =-i \epsilon_{\a\b} \tilde F_{\da\db} +i \epsilon_{\da\db} F_{\a\b}  \,,
\end{align}
which are related by complex conjugation $\left[ F_{\a\b} \right]^* = \tilde F_{\da\db}$.
Then both the Yang-Mills part of the Lagrangian \p{Lagr} and the pseudo-scalar $F^{\mu\nu} \tilde F_{\mu\nu}$  split into chiral and anti-chiral pieces, 
\begin{align}
F_{\mu \nu} F^{\mu \nu} = \frac{1}{2} \left( F_{\a\b} F^{\a\b} + \tilde F_{\da\db} \tilde F^{\da\db} \right)\,, \\
F_{\mu \nu} \tilde F^{\mu \nu} = \frac{i}{2} \left( F_{\a\b} F^{\a\b} - \tilde F_{\da\db} \tilde F^{\da\db} \right)\,,
\end{align}
related by complex conjugation. 
Thus, it is natural to introduce the chiral and anti-chiral forms of the YM Lagrangian 
\begin{align}
L = -\frac{1}{2}\tr (F_{\a\b})^2 \;, \quad \bar L = -\frac{1}{2}\tr (F_{\da\db})^2\,,
 \label{chirYM}
\end{align}
along with the real YM Lagrangian
\begin{align}
{\cal L} = -\frac{1}{2}\tr\, F_{\mu \nu} F^{\mu \nu} \,. \label{realYM}
\end{align}
The chiral, anti-chiral and real Lagrangians differ by total derivatives, so they produce the same action $S_{\rm YM}= \int d^4x \cL = \int d^4x L = \int d^4x \bar L $. We see that the imaginary part of the correlator with the chiral Lagrangian insertion 
\begin{align}
G^{(\rm chir)}_{\mu\nu}(x_1,x_2,x_3) =\vev{J_{\mu}(x_1)\, J_{\nu}(x_2) \, L(x_3) }_{\rm Born} \, \label{corrJJLchir}
\end{align}
yields the correlator \p{corrJJTh} involving the pseudo-scalar (or parity odd) topological term $\Theta$,
\begin{align}
G^{(\rm chir)}_{\mu\nu} - \left[ G^{(\rm chir)}_{\mu\nu} \right]^* = i G^{(\rm odd)}_{\mu\nu}\,. \label{odd}
\end{align}
Thus, it will be enough to evaluate the complex correlator \p{corrJJLchir}. The Lagrangian insertions method is a powerful tool for generating multi-loop integrands of correlation functions in supersymmetric theories \cite{Eden:2000mv,Eden:2010zz,Eden:2011ku,Eden:2012tu} and in massless QCD \cite{Chicherin:2020azt}.
Let us note that by taking the real part of \p{corrJJLchir} we find the real YM Lagrangian insertion \p{realYM} in the correlator of two currents,
\begin{align}
\frac{1}{2} \left( G^{(\rm chir)}_{\mu\nu} + \left[ G^{(\rm chir)}_{\mu\nu} \right]^* \right) = \vev{J_{\mu}(x_1)\, J_{\nu}(x_2) \, {\cal L}(x_3)  }_{\rm Born} \equiv G^{(\rm even)}_{\mu\nu}(x_1,x_2,x_3)\,. \label{even}
\end{align}

Now we turn to the calculation of the Feynman diagrams in Fig.~\ref{3pt} but with the chiral operator $L$ at point $x_3$. Even at the lowest perturbative level we have to carry out nontrivial space-time integrations.

\subsection{Factorizable diagram: T-block $\times$ T-block}

The leftmost diagram in Fig.~\ref{3pt} factorizes in a product of two T-blocks  depicted on the lhs of Fig.~\ref{FigPsipsiF2}. The T-block depends on three external points and involves one space-time integration vertex (here $(\b\g)$ denotes weighted symmetrization),
\begin{align}
&\vev{\psi_\a(x_1) \bar\psi_\da(x_2) F^{a}_{\b\g}(x_3)}_{g}  = \frac{2i g\, T^a}{(2\pi)^6}
\int d^4 x_0 \, \pa_{\a\dot\delta}\frac{1}{x_{10}^2}  \pa_{(\beta\da} \frac{1}{x_{20}^2}
\pa^{\dot\delta}_{\gamma)}\frac{1}{x_{30}^2} \notag\\
&= \frac{4 g\, T^a}{(2\pi)^4}\left\{ \frac{(x_{12})_{\a\da}}{x^4_{12}} \frac{(x_{31} \tx_{32})_{(\b\g)}}{x^2_{13} x^2_{23}} + \frac{(x_{12} \tx_{23})_{\a(\b}  (x_{32})_{\g)\da}}{x^2_{12} x^2_{13} x^4_{23}} \right\}\,.
\label{T-block}
\end{align}
 The rationality of the T-block \p{T-block} follows from the `star-triangle' identity 
\begin{align}
(\pa_1)_{\a\da} (\pa_2)^{\da\b} \int  \frac{d^4 x_0}{x^2_{10} x^2_{20} x^2_{30} } = 4\pi^2 i\frac{(x_{13} \tx_{32})_{\a}{}^{\b}}{x^2_{12} x^2_{13} x^2_{23}} \,, \label{str-trng}
\end{align}
which is a consequence of the conformal covariance of this three-point integral (a Yukawa vertex).

\subsection{Chiral insertion into the fermion propagator} 

\begin{figure}
\begin{center}
\begin{tabular}{cc}
\includegraphics[width=5.5cm]{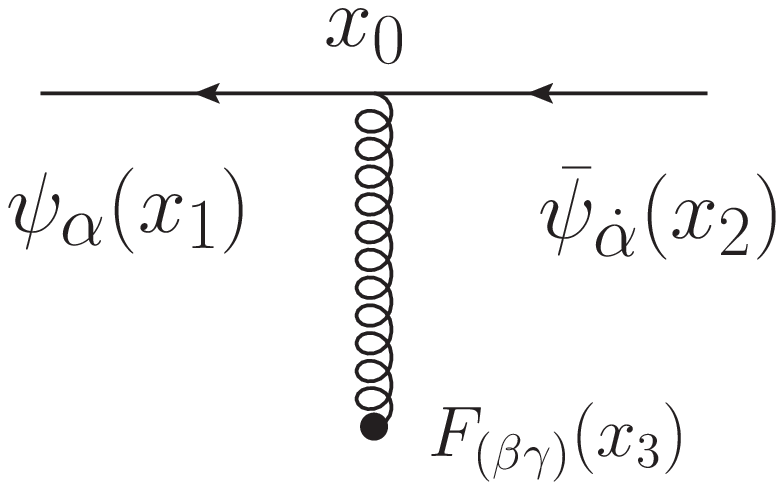} \qquad
&
\qquad
\includegraphics[width=7cm]{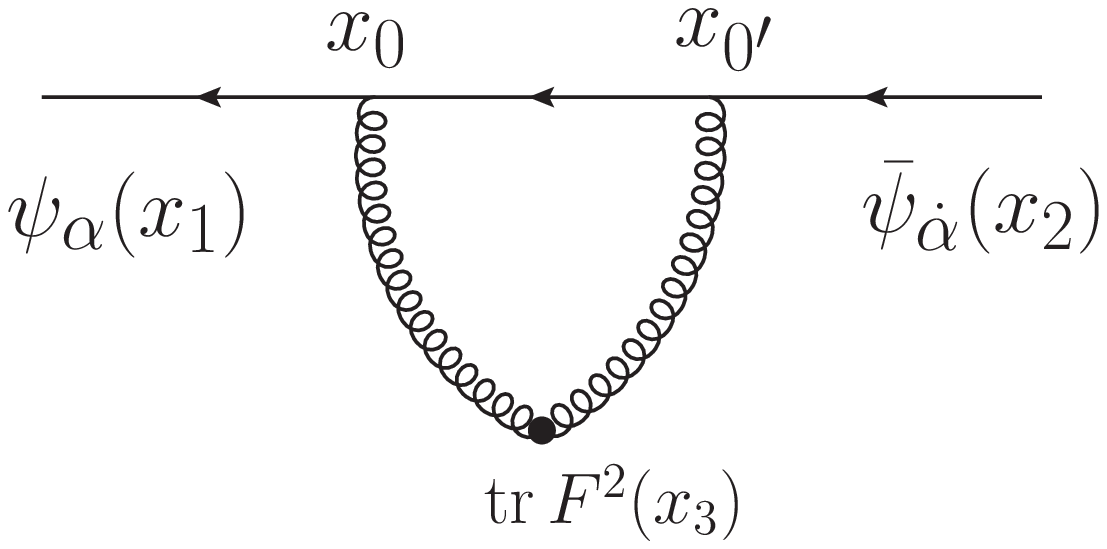}
\end{tabular}
\end{center}
\caption{Building blocks of the correlator Feynman diagrams.}\label{FigPsipsiF2}
\end{figure}

The remaining two Feynman diagrams in Fig.~\ref{3pt} involve a genuine two-loop Feynman integral depicted on the rhs of Fig.~\ref{FigPsipsiF2}. It can be interpreted as the insertion of the chiral YM Lagrangian $L(x_3)$ in the fermion propagator $\vev{\psi(x_1) \bar\psi(x_2)}$,  
\begin{align}
&\Pi_{\a\da}(x_1,x_2,x_3) \equiv \vev{\psi_\a(x_1) \bar\psi_\da(x_2)\, \tr \, (F_{\b\g}^2)(x_3)}_{g^2} \notag\\
& =  -\frac{4i g^2 C_F}{(2\pi)^{10}} \int d^4 x_0 d^4 x_{0'}\, \pa_{\a\db} \frac{1}{x_{10}^2} \pa_{(\b\da} \frac{1}{x_{20'}^2} \pa_{\gamma) \dot\gamma} \frac{1}{x_{30'}^2} \pa^{\dot\gamma(\gamma} \frac{1}{x_{00'}^2} \pa^{\db \beta)} \frac{1}{x_{30}^2}\,.\label{eq1.15}
\end{align}
We carry out one of the integrations, e.g. at point $x_{0'}$, in four dimensions  by means of the star-triangle identity \p{str-trng}. Then after simplifying the resulting integrand we arrive at
\begin{align}
\Pi_{\a\da}(x_1,x_2,x_3) = \frac{g^2 C_F}{(2\pi)^8}(\pa_{1})_{\a\db}(\pa_2)_{\b\da} & \biggl[ -3 \pa_{3}^{\db\b} \int \frac{d^D x_0}{x_{10}^2 x_{20}^2 x_{30}^4}  \notag\\ 
&+ \frac{(x_{23})_{\gamma\dot\gamma}}{2 x_{23}^2} \left( \pa_3^{\dot\gamma\beta} \pa_3^{\db\gamma} + \pa_3^{\dot\gamma\gamma} \pa_3^{\db\b} \right) \int \frac{d^D x_0}{x_{10}^2 x_{20}^2 x_{30}^2} \biggr] \label{propcorDimReg}
\end{align} 
where we temporarily introduced dimensional regularization with $D=4-2\ep$. We expect the regular part of the Feynman integral \p{eq1.15} to be finite (see below).  Keeping $D=4$, in the first term we can write $\pa_3=-(\pa_1+\pa_2)$ and $(\pa_{1})_{\a\db}(\pa_2)_{\b\da} (\pa_{3})^{\db\b} = -\Box_1 (\pa_2)_{\a\da} -\Box_2 (\pa_1)_{\a\da} $. Taking into account that $\Box_{1} \frac{1}{x_{10}^{2}} = 4i\pi^2 \delta^{(4)}(x_{10})$ we lift the integration by these delta functions.  The second integral in \p{propcorDimReg} is done again by the star-triangle identity \p{str-trng} with respect to points 1 and 3. Thus, the two space-time integrations in \p{propcorDimReg} result in rational functions and we obtain 
\begin{align} 
\Pi_{\a\da}(x_1,x_2,x_3) =- \frac{4 i g^2 C_F}{(2\pi)^{6}} \frac{1}{x_{13}^4 x_{23}^4}\left\{ (x_{23})_{\a\da} + \frac{(x_{12})_{\a\da}}{2 x_{12}^4} \left[ -x_{13}^4 - x_{23}^4 + 2 x_{13}^2 x_{23}^2 + 4 x_{12}^2 x_{13}^2\right]\right\}.
\label{propCorr}
\end{align}
We note that by hitting the Feynman integral \p{eq1.15} with the Dirac operator $\pa_1^{\db\a}$, we lift the space-time integration at one of the interaction vertices and the remaining space-time integration is again reduced to the star-triangle identity \p{str-trng}. Thus we derive a differential equation of the form $\pa^{\da\a}_1 \Pi_{\a\da} =$ `known rational function'. One can easily check that \p{propCorr} satisfies this DE. An analogous DE with respect to $x_2$ is satisfied as well.

The regular part of \p{eq1.15} is free from UV divergences, but they appear in the form of contact terms omitted in \p{propCorr}.
Indeed, the only possible source of divergences in \p{propcorDimReg} is the integration region $x_0 \sim x_3$ in the first term. In order to extract the $\ep$-pole of the diagram, we expand the singular $D-$dimensional distribution 
\begin{align}
\frac{1}{x^4} \to \frac{i\pi^2}{\ep} \delta^{(4)}(x) + {\cal O}(\ep^0) \label{delta}
\end{align}
with $x = x_{30}$ in eq. \p{propcorDimReg}, and find
\begin{align}
\Pi_{\a\da} (x_1,x_2,x_3) 
= \frac{3 g^2 C_F}{2(2\pi)^4}\frac{1}{\epsilon} \left( -\delta(x_{13}) +  \delta(x_{23})  \right) \frac{(x_{12})_{\a\da}}{x_{12}^4}+ {\cal O}(\ep^0) \,. \label{contact}
\end{align}

Concluding this subsection we mention that the integration over the insertion point $x_3$ gives the one-loop correction to the fermion propagator: 
\begin{align}\label{}
\lim_{\ep\to0} \int d^{D}x_3\, \Pi_{\a\da}(x_1,x_2,x_3) =  -\frac{3 g^2 C_F}{(2\pi)^{4}}  \frac{(x_{12})_{\a\da}}{x^4_{12}}\,.
\end{align}
Note that the poles in the contact terms \p{contact} cancel out and the correction is finite. 
This result differs from the familiar infinite propagator correction  in the Feynman gauge or vanishing correction   in the Landau gauge (see, e.g.,  \cite{Grozin:2005yg}). The explanation is that this quantity is not only gauge but also scheme dependent;  our Lagrangian insertion procedure constitutess a different scheme.

\subsection{Born-level three-point correlator}
\label{secBL3pC}

Using the expressions for the building blocks \p{T-block} and \p{propCorr} of the Feynman diagrams in Fig.~\ref{3pt}, we find the Born-level correlator \p{corrJJLchir} of two currents and the chiral YM Lagrangian $L$ \p{chirYM},   
\begin{align}\label{JJLdiagr}
G^{(\rm chir)}_{\a\da\b\db}(x_1,x_2,x_3) =  -\frac{4g^2 C_F}{(2\pi)^8} & \biggl[ \frac{(x_{12})_{\a\db} (x_{21})_{\b\da}}{x_{12}^4 x_{13}^4 x_{23}^4} + \frac{2(x_{13} \tilde x_{32})_{\a\b} (\tilde x_{13} x_{32})_{\da\db}}{x_{12}^2 x_{13}^6 x_{23}^6} \notag\\
&+ \frac{3\ep_{\alpha\beta}(\tilde x_{13} x_{32})_{\dot\alpha\dot\beta}}{x_{12}^4 x_{13}^4 x_{23}^4} + \frac{3\ep_{\dot\alpha\dot\beta}(x_{13} \tilde x_{32})_{\alpha\beta}}{x_{12}^4 x_{13}^4 x_{23}^4} \biggr].
\end{align}
{One can easily check that the correlator satisfies current conservation at points $x_1$ and $x_2$.}

The Lorentz spinor notation makes it obvious that the first line in \p{JJLdiagr} is real and conformally covariant, whereas the second line is imaginary and it breaks the conformal symmetry, see \p{CC}, \p{CT}.\footnote{The two-component Levi-Chivita tensors in the second line  are not covariant under conformal inversion.  } According to eq. \p{odd} the latter corresponds to the correlator with the pseudo-scalar insertion $\Theta$ \p{corrJJTh},\footnote{Here we are still using the generic notation $J$ for a current made from a single spinor. If specified to vector or axial currents as in \p{2.5}, \p{2.6}, it is easy to see that the correlator vanishes for the parity-even combination $\vev {V A \Theta}$ and has the form \p{LeviChivita} for the parity-odd combinations $\vev{VV\Theta}$ and $\vev{AA\Theta}$.  }
\begin{align}
G^{(\rm odd)}_{\mu\nu}(x_1,x_2,x_3) \equiv \vev{J_\mu(x_1)\,J_\nu(x_2)\, \Theta(x_3)}_{\rm Born} = -\frac{24 g^2 C_F}{(2\pi)^8}  \frac{\ep_{\mu\nu\la\rho} x^\la_{13} x^\rho_{23}}{x_{12}^4 x_{13}^4 x_{23}^4} \,.\label{LeviChivita}
\end{align}
One can easily see that  \p{LeviChivita} satisfies the current conservation conditions  $\pa^{\mu}_1 G^{(\rm odd)}_{\mu\nu} = \pa^{\nu}_2 G^{(\rm odd)}_{\mu\nu} = 0$ (up to contact terms). It is a finite\footnote{Up to  contact terms, see  \p{contact}.} rational function, which is not conformal. Thus the conformal symmetry is broken already at the lowest perturbative level.

On the contrary, the real part of \p{JJLdiagr} yields the real Lagrangian insertion \p{even} which is explicitly conformal,
\begin{align}
G^{(\rm even)}_{\mu\nu}(x_1,x_2,x_3)  = \frac{2g^2 C_F}{(2\pi)^8}\frac{1}{x_{13}^4 x_{23}^4}\left[ \frac{1}{x_{12}^2} I_{\mu\nu}(x_{12}) - 4 Z_\mu(x_1|x_2,x_3) Z_\nu (x_2|x_1,x_3) \right]. 
\end{align}
Here we employ the familiar conformal tensors (see, e.g., \cite{Schreier:1971um,Erdmenger:1996yc})
\begin{align}\label{2.27}
I^{\mu\nu}(x_{12}) = \eta^{\mu\nu} - \frac{2 x_{12}^\mu x_{12}^\nu}{x_{12}^2} \,,\qquad
Z_\mu (x_1|x_2,x_3) = \frac{x_{12}^\mu}{x_{12}^2} - \frac{x_{13}^\mu}{x_{13}^2} \,.
\end{align}

\section{Anomalous conformal Ward identity}
\label{secAnom}

In the previous Section we verified by an explicit Feynman diagram calculation that the three-point correlator \p{corrJJTh} involving the pseudo-scalar topological term $\Theta$ is not conformal already at the Born level. The interaction vertices, operator vertices and propagators respect the conformal symmetry, since they originate from the conformal Lagrangian \p{Lagr}. The space-time integrations do not introduce divergences. Thus, we would naively expect unbroken conformal symmetry at this perturbative level. Now we explore in more details the mechanism leading to the conformal symmetry breaking and calculate the corresponding anomaly.

The  breakdown of conformal symmetry could only come from hidden singularities in the Feynman integrals. They are revealed when we perform a  conformal variation under the sign of the integral in dimensional regularization with $D =4-2\ep$ \cite{Braun:2003rp,Drummond:2007au}.  More precisely, we modify only the dimension of the measure but not that of the fields and coupling constant. Due to the mismatch of the conformal weights of the $D$-dimensional measure and the four-dimensional Lagrangian, we find the conformal variation (see \p{Kboost}) of the action  
\begin{align}
\mathbb{K}^\la \int d^D x\, L_{\rm QCD}(x) = 2i(D- \Delta_{L}) \int d^D x \, x^\la \, L_{\rm QCD}(x) \,,
\end{align}
where $\Delta_L =4$ is the conformal weight of the Lagrangian \p{Lagr}. The variation is of order ${\cal O}(\ep)$. Thus, the conformal variation of a Feynman diagram amounts to inserting $\ep x^{\la}$ in the interaction vertices and promoting the space-time integrations to $D$ dimensions. If this modification leads to a UV (i.e., short distance) divergence in the space-time integrals, i.e. an $\ep$-pole, then the variation of the Feynman diagram is of $O(\ep^0)$ as $\ep \to 0$ and the conformal symmetry is anomalous. The corresponding anomalous conformal Ward identity takes the form  
\begin{align}
&\mathbb{K}^{\la} \,\vev{J_{\mu}(x_1) \, J_\nu(x_2)\, L(x_3)}_{\rm Born} \notag\\
&= 4\lim_{\ep\to 0} \ep \int d^D x_0\, x_0^\la  \vev{J_{\mu}(x_1) \, J_\nu(x_2)\, L(x_3)\, L_{\rm QCD}(x_0)}_{\rm Born} \,. \label{CAWI}
\end{align}

In order to evaluate the rhs of the Ward identity we inspect the Feynman diagrams in Fig.~\ref{3pt} and successively insert an extra factor $x_0$ in the interaction vertices. The insertion of $x_0$ in the T-block integrations \p{T-block} does not create an $\ep$-pole, so we can ignore the first diagram. However, the insertion of $x_0$ or $x_0'$ into the propagator correction type diagrams in Fig.~\ref{3pt} does create a UV divergence. Thus, only the Feynman integral $\Pi_{\a\da}$ \p{eq1.15} is responsible for the conformal anomaly. There are two contributions to the conformal anomaly: (i) insertion of $x_0^\la (\sigma_\la)^{\dot\gamma\gamma}=x_0^{\dot\gamma\gamma}$ in the left interaction vertex in Fig.~\ref{FigPsipsiF2}; (ii) insertion of $x_{0'}^{\dot\gamma\gamma}$ in the right interaction vertex in Fig.~\ref{FigPsipsiF2}. In the following we ignore the regular part of the conformal variation, which is guaranteed to cancel in the sum of all diagrams, and keep only the anomalous part, which contributes to the rhs of \p{CAWI}. In order to evaluate the first anomalous contribution we simplify the fermion propagator correction $\Pi_{\a\da}$ \p{eq1.15} as in \p{propcorDimReg} and do the insertion of $x_0$ only in the first term in \p{propcorDimReg}. The source of the UV pole is the singular distribution $1/x^4_{30}$  (see 
\p{delta}). We find    
\begin{align}\label{104}
\left[ \mathbb{K}_{\gamma\dot\gamma} \Pi_{\a\da} \right]_{\rm anom\,1} &=-\lim_{\epsilon \to 0} 4i\ep \,
(\pa_{1})_{\a\db}(\pa_2)_{\beta\da} \pa_3^{\db\b} \frac{g^2 C_F }{(2\pi)^8}\int d^D x_0 \frac{ -3(x_{0})_{\gamma\dot\gamma}}{x_{10}^2 x_{20}^2 }\left[ \frac{i\pi^2}{\ep} \delta^{(4)}(x_{30}) + {\cal O}(\ep^0) \right] \notag\\
&= -\frac{24 g^2 C_F}{(2\pi)^6}  \frac{(x_{13})_{\a\dot\gamma}(x_{23})_{\gamma\da}}{x_{13}^4 x_{23}^4} \,.
\end{align}
We recall that without the insertion of $(x_{0})_{\gamma\dot\gamma}$ the integral produces the contact term \p{contact}, which indicates the hidden singularity.

The second anomaly contribution is obtained by first integrating out the $x_0$ vertex  in \p{eq1.15} by means of the star-triangle identity \p{str-trng},
and then rewriting the resulting integral with dimensional regularization  (we relabel $x_{0'}$ to $x_0$):
\begin{align}
\Pi_{\a\da} = -\frac{2g^2 C_F}{(2\pi)^8} \int d^D x_0 \biggl[ 
\frac{3 (x_{10})_{\a\db}}{x_{13}^2 x_{10}^2} \, \pa_{\b \da} x_{20}^{-2} \,
\pa^{\db\b} x_{30}^{-4}
+ \frac{(x_{13})^{\db(\delta}}{x_{13}^2} \left( \pa^{\b)\dot\delta} \frac{(x_{10})_{\a\db}}{x_{10}^2}\right) \pa_{\b\da} x_{20}^{-2}\pa_{\delta\dot\delta} x_{30}^{-4} 
\biggr].
\end{align}
One can easily verify that the integral is finite (up to contact terms). Inserting $x_0^{\dot\gamma\gamma}$ into it and extracting the $\ep$-pole by means of \p{delta}, we find the second anomaly contribution
\begin{align}
\left[ \mathbb{K}_{\gamma\dot\gamma} \Pi_{\a\da} \right]_{\rm anom\,2}  = -\frac{24 g^2 C_F}{(2\pi)^6}\frac{(x_{13})_{\a\dot\gamma}(x_{23})_{\gamma\da}}{x_{13}^4 x_{23}^4} \,. \label{anom2}
\end{align} 
Collecting the anomalous contributions \p{104} and \p{anom2}, we derive the conformal anomaly of $G^{(\rm chir)}$ written in spinor notation \p{JJLdiagr},
\begin{align}\label{3.6} 
\mathbb{K}^{\dot\gamma\gamma}\, (G^{(\rm chir)})^{\a\da\b\db} = \frac{24 i g^2 C_F}{(2\pi)^8} \frac{1}{x_{12}^4 x_{13}^4 x_{23}^4} \left[ x_{32}^{\dot\gamma\beta} x_{13}^{\da\gamma} x_{21}^{\db\a} + x_{31}^{\dot\gamma\a}  x_{23}^{\db\gamma} x_{12}^{\da\b} \right]\,.
\end{align}

We recall that this anomaly originates from the insertion of the chiral Lagrangian \p{corrJJLchir} into the two-point function of two real vector currents. Repeating the whole procedure with the anti-chiral Lagrangian, we get the { complex conjugat }of \p{3.6}. In this way we see that the anomalies cancel for the real (parity even) Lagrangian insertion \p{even}, while they add up for the imaginary (parity odd) insertion \p{LeviChivita}. Converting \p{3.6} to the vector notation, we find the conformal anomaly of  \p{LeviChivita}
\begin{align}\label{1011}
&\mathbb{K}_\la\, \vev{J_{\mu}(x_1)\, J_{\nu}(x_2)\, \Theta(x_3)}_{\rm Born} = \frac{12 i g^2 C_F}{(2\pi)^8}\frac{1}{x_{12}^4 x_{13}^4 x_{23}^4}\, \biggl[(x^2_{13} +x^2_{23} - x_{12}^2) \ep_{\mu\nu\la\rho}  x^\rho_{12} \notag\\ 
&+ 2 (x_{13}+x_{23})_\la \ep_{\mu\nu\kappa\tau} x^\kappa_{13} x^\tau_{23} - 2 (x_{12})_\mu \epsilon_{\nu\la\kappa\tau} x_{13}^\kappa x_{23}^\tau  -2 (x_{12})_\nu \epsilon_{\mu\la\kappa\tau} x_{13}^\kappa x_{23}^\tau  \biggr].
\end{align}
We have checked by an explicit calculation that the rhs of \p{LeviChivita} does indeed verify the anomalous conformal Ward identity \p{1011}. 

In summary, we have shown that the conformal anomaly is due to hidden singularities in the Feynman integrals. The integral itself is finite (up to contact terms) but a conformal transformation changes the balance of powers and causes an UV divergence. Its effect is a non-vanishing conformal variation.

\section{Conformal anomaly of the Yukawa operators}

The subtle effect of conformal symmetry breaking by pseudo-scalars operators is not limited to gauge theories and to the topological term $\Q=\tr\,F_{\mu\nu}\tilde F^{\mu\nu}$. In this Section we show another, even simpler example of conformal symmetry breaking by the chiral Yukawa vertex in four dimensions. 

Let us consider a massless complex scalars $\varphi$ coupled to a massless (anti)chiral fermion $\psi (\bar\psi)$  described by the real Lagrangian 
\begin{align}
L_{\rm Yuk} = \pa^{\mu} \bar \varphi \,\pa_{\mu} \varphi + \frac{i}{2} \psi^\a \overset\leftrightarrow{\pa}_{\a\da} \bar\psi^\da + g \,\varphi\, \psi^\a \psi_\a + g\, \bar\varphi \, \bar\psi_\da \bar\psi^\da \,. \label{YukLagr}
\end{align}
The Lagrangian is classically conformal. Global $U(1)$ transformations are generated by the classically conserved current 
\begin{align}
J_{\a\da} = \psi_{\a} \bar\psi_{\da} - \frac{i}{2} \varphi \,\pa_{\a\da} \bar\varphi + \frac{i}{2} \bar\varphi\, \pa_{\a\da} \varphi\,, \qquad \pa^{\da\a} J_{\a\da} = 0 \,. \label{curr}
\end{align}  
The Lagrangian comprises chiral and antichiral Yukawa vertices,
\begin{align}
Y = g \,\varphi\, \psi^\a \psi_\a \;,\qquad
\overline{Y} = g\, \bar\varphi \, \bar\psi_\da \bar\psi^\da \,, \label{YYbar}
\end{align}
which are related by complex conjugation. We can form a real (it appears in the Lagrangian \p{YukLagr}) and an imaginary combinations out of them,
\begin{align}
\widehat{Y} \equiv Y + \overline{Y} \;,\qquad
\widetilde Y \equiv Y - \overline{Y} = g \,\varphi\, \psi^\a \psi_\a -  g\, \bar\varphi \, \bar\psi_\da \bar\psi^\da \,, \label{YY}
\end{align}  
which are parity even and odd, respectively.\footnote{The parity properties become more transparent if we use a four-component Majorana spinor $\Psi=(\psi, \bar\psi)$. The complex scalar field $\varphi=S+iP$ comprises  a scalar $S$ and a pseudo-scalar $P$. Then the two combinations in \p{YY} become $\widehat Y= ig\bar\Psi(S-\g_5 P)\Psi$ and $\widetilde{Y}=g\bar\Psi(P+\g_5 S)\Psi$. Also, the current \p{curr} is an axial vector.}

\begin{figure}
\begin{center}
\begin{tabular}{cc}
\includegraphics[width=7cm]{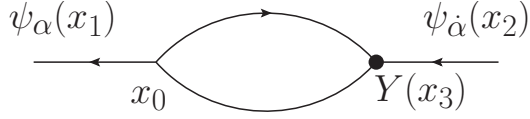} \qquad
\end{tabular}
\end{center}
\caption{Insertion of the chiral $Y$ in the fermion propagator.}\label{FigPsiPsiY}
\end{figure}

Now we consider the three-point correlator of two currents \p{curr} and a chiral Yukawa operator,
\begin{align}
\vev{J_{\mu}(x_1)\, J_{\nu}(x_2)\, Y(x_3)}_{\rm Born} \,. \label{JJYL}
\end{align}
In the Born approximation it is a rational function  of order ${\cal O}(g^2)$. There are several Feynman diagrams contributing to \p{JJYL}. We restrict our attention to the diagrams with the insertion of $Y$ into the fermion propagator (see Fig.~\ref{FigPsiPsiY}). The remaining diagrams contributing to \p{JJYL} are conformal\footnote{This is in contrast with  a gauge theory where the individual Feynman diagrams are gauge dependent and hence not conformal.} and can be easily evaluated by means of the star-triangle identity \p{str-trng}. We find
\begin{align}
& \Pi_{\a\da} \equiv \vev{\psi_{\a}(x_1)\, \bar\psi_{\da}(x_2)\,Y(x_3)}_{g^2} \notag\\
& = -\frac{4g^2}{(2\pi)^{8}} \pa_{\b\da} x_{32}^{-2} \int d^4 x_0 \, x_{30}^{-2}\, \pa^{\db\b} x_{30}^{-2} \, \pa_{\a\db} x_{10}^{-2} = \frac{4g^2}{(2\pi)^6} \frac{(x_{23})_{\a\da}}{x_{13}^4 x_{23}^4}\,, \label{propY}
\end{align}
where we used the identity $x_{30}^{-2}\, \pa^{\db\b} x_{30}^{-2} = \frac{1}{2}\pa^{\db\b} x_{30}^{-4}$ and integrated by parts, thus producing $\Box x_{10}^{-2} = 4 i \pi^2 \delta^{(4)}(x_{10})$.
The complex conjugate gives the antichiral insertion  $\overline{Y}$ into the fermion propagator,
\begin{align}\label{4.7}
\vev{\psi_{\a}(x_1)\, \bar\psi_{\da}(x_2)\,\overline{Y}(x_3)}_{g^2} =  -\frac{4i g^2}{(2\pi)^6} \frac{(x_{13})_{\a\da}}{x_{13}^4 x_{23}^4}\,.
\end{align} 
Combining the two insertions we find that the real one $\widehat{Y}$ is conformal, 
\begin{align}\label{4.8} 
\vev{\psi_{\a}(x_1)\, \bar\psi_{\da}(x_2)\,\widehat{Y}(x_3)}_{g^2}= -\frac{4ig^2}{(2\pi)^6} \frac{(x_{12})_{\a\da}}{x_{13}^4 x_{23}^4}\,,
\end{align}
while the imaginary one $\widetilde Y$ is not. Thus, the Born-level three-point correlator involving the parity-even operator $\widehat{Y}$ is conformal,  
\begin{align}
\mathbb{K}_\la \,\vev{J_{\mu}(x_1)\, J_{\nu}(x_2)\, \widehat{Y}(x_3)}_{\rm Born} = 0 \,, \label{confJJYhat}
\end{align}
while the correlator $\vev{J_\mu J_\nu \widetilde{Y}}$ involving the pseudo-scalar $\widetilde Y$ is not conformal. 

The mechanism of conformal symmetry breaking is the same as in the gauge sector in Section~\ref{secAnom}: A singular distribution $1/x^4$ in the integral at the interaction vertex in Fig.~\ref{FigPsiPsiY} causes a pole which results in an anomaly term in the conformal variation.  For the chiral insertion $Y$, the analog of eq.~\p{CAWI} takes the following form
\begin{align}
& \mathbb{K}^\la \,\vev{J_{\mu}(x_1)\, J_{\nu}(x_2)\, Y (x_3)}_{\rm Born} = \notag\\
&= 4\lim_{\ep\to 0} \ep \int d^D x_0\, x_0^\la  \vev{J_{\mu}(x_1) \, J_\nu(x_2)\, Y(x_3)\, L_{\rm Yuk}(x_0)}_{\rm Born} \,. \label{CAWI2}
\end{align}
Only the propagator correction type diagram contributes to the anomaly. The remaining diagrams are conformal and can be ignored. Then according to \p{CAWI2} the anomalous part of the conformal variation of $\Pi_{\a\da}$ \p{propY} is given by
\begin{align}
\left[ \mathbb{K}^{\dot\gamma\gamma} \, \Pi_{\a\da} \right]_{\rm anom} =  \frac{16i g^2}{(2\pi)^8}\lim_{\ep \to 0} \ep \int d^D x_0 \, x_{0}^{\dot\gamma\gamma} x_{30}^{-2} \pa^{\db\b} x_{30}^{-2} \pa_{\a\db} x_{10}^{-2} \pa_{\b\da} x_{32}^{-2}\,.
\end{align}
In order to extract the pole we proceed as in \p{propY} and replace $x_{30}^{-4}$ by its pole part \p{delta},
\begin{align}
\left[ \mathbb{K}^{\dot\gamma\gamma} \, \Pi_{\a\da} \right]_{\rm anom} & = \frac{8ig^2}{(2\pi)^8}\lim_{\ep \to 0} \ep \int d^4 x_0\, \pa^{\db\b} \left[  x_{0}^{\dot\gamma\gamma} \pa_{\a\db} x_{10}^{-2} \right] \frac{i\pi^2}{\ep} \delta^{(4)}(x_{30})\,  \pa_{\b\da} x_{32}^{-2} \notag\\
& = -\frac{16g^2}{(2\pi)^6}\frac{(x_{13})_{\a}^{\dot\gamma} (x_{32})_{\da}^{\gamma}}{x_{13}^4 x_{23}^4} \,,
\end{align}
where we omitted the  contact term. Then the anomalous conformal Ward identity for the three-point correlator with the chiral $Y$ takes the following form 
\begin{align}
&\mathbb{K}^{\dot\gamma\gamma} \, \vev{J^{\a\da}(x_1)\, J^{\b\db}(x_2)\, Y (x_3)}_{\rm Born} \notag\\
&= \frac{16 i g^2}{(2\pi)^8} \frac{1}{x_{12}^4 x_{13}^4 x_{23}^4} \left[ x_{21}^{\db\a} x_{13}^{\da\gamma} x_{32}^{\dot\gamma\b} + x_{12}^{\da\b} x_{31}^{\dot\gamma\a} x_{23}^{\db\gamma}  \right]\,. \label{JJYanom}
\end{align}
The anomaly of $\vev{J_{\mu} J_{\nu} \overline{Y} }$ has the opposite sign, so the two anomalies cancel in the real combination $\widehat{Y}$ and double in the pseudo-scalar combination $\widetilde Y$. Remarkably, the expression for the anomaly \p{JJYanom} is identical (up to normalization) with the anomaly in the gauge sector \p{3.6}.

\section{Conformal anomaly cancellation in $\cN=4$ SYM}
\label{secSusy}

So far we have observed that the pseudo-scalar operators $\Theta$ \p{Theta} and $\widetilde Y$ \p{YY} from the gauge and Yukawa sectors, respectively, break the conformal symmetry at the lowest perturbative level. Now we are going to show that supersymmetry helps to restore conformal symmetry.

We consider the maximally supersymmetric  $\cN=4$ Yang-Mills theory. It comprises: gauge bosons, pseudo-real scalars $\phi_{AB}=\frac1{2} \ep_{ABCD} \bar\phi^{CD}$ in the antisymmetric ${\bf 6}$ representation of the R-symmetry $SU(4)$, (anti)chiral fermions $\psi^A_{\a}$ and $\bar\psi_A^{\da}$ in the (anti)fundamental  representation of $SU(4)$. All the fields are massless and they transform in the adjoint representation of the color group $SU(N_c)$. The theory is conformal at the quantum level, i.e. $\beta(g) = 0$ to all orders in the coupling. The gauge-invariant composite operators form multiplets of supersymmetry. The stress-tensor multiplet is of particular interest, being a protected half-BPS multiplet. The operators in the multiplet are not renormalized and keep their canonical dimensions. The multiplet contains all the conserved currents of the theory, as well as the chiral on-shell Lagrangian
\begin{align}\label{A.11}
L_{{\cal N}  = 4}
& = \tr \left\{- \frac12  F_{\alpha\beta}F^{\alpha\beta}  + {\sqrt{2}}  g \psi^{\alpha A} [\phi_{AB},\psi_\alpha^B] - \frac18 g^2 [\phi^{AB},\phi^{CD}][\phi_{AB},\phi_{CD}] \right\} 
\end{align}  
and its anti-chiral conjugate.
The real part of $L_{{\cal N} = 4}$ is a conformal parity-even scalar. Its  imaginary part is a parity-odd scalar including the gauge sector pseudo-scalar $\Theta$ \p{Theta} and the pseudo-scalar Yukawa term $\widetilde{Y}$ \p{YY},
\begin{align}\label{5.2} 
{\rm Im}\,L_{{\cal N}  = 4} =  \tr \left\{ \frac{i}{4} \left( F_{\alpha\beta}F^{\alpha\beta}- \bar F_{\dot\alpha\dot\beta}\bar F^{\dot\alpha\dot\beta} \right) - \frac{ig}{\sqrt{2}}  \psi^{\alpha A} [\phi_{AB},\psi_\alpha^B] - \frac{ig}{\sqrt{2}}  \bar\psi_{\da A} [\phi^{AB},\bar\psi^\da_B]  \right\}\,.
\end{align}
Using the equations of motion we can rewrite the Yukawa terms as a total derivative,
\begin{align}\label{A.13}
{\rm Im}\,L_{{\cal N}  = 4}\bigr|_{\rm EOM} =   \frac{i}{4} \tr\left( F_{\alpha\beta}F^{\alpha\beta}- \bar F_{\dot\alpha\dot\beta}\bar F^{\dot\alpha\dot\beta} \right) -  \partial^{\dot\alpha \alpha} \tr (\bar\psi_{\dot\alpha A} \psi_\alpha^A) \,,
\end{align}
therefore the action $S_{{\cal N}  = 4} = \int d^4x\, L_{{\cal N}  = 4} = \int d^4x\, \bar L_{{\cal N}  = 4}$ is real. The fermion term in \p{A.13} is the divergence of the $U(1)$ axial current which completes the R-symmetry $SU(4)$ to $U(4)$ in the free theory. In the interacting theory this $U(1)$ symmetry is broken by the Yukawa term because the real scalars $\phi$ have no $U(1)$ charge. 

The main point we would like to make in this section is the cancellation of the conformal anomalies of $\Theta$ and $\widetilde{Y}$ in the particular combination that appears in \p{5.2}. The same applies to the complex chiral Lagrangian \p{A.11}. We conclude that $L_{{\cal N}  = 4}$  is a conformal primary operator, as expected from a member of the super-conformal multiplet of the energy-momentum tensor.

\section{Axial anomaly and conformal anomaly}
\label{secChirAnom}

In the previous sections we have shown that the topological term $\Q$ is not a conformal primary operator. This manifests itself in the fact that its correlation functions with other, conformal operators are not conformal. Here we give an alternative interpretation of this phenomenon, based on the relationship between $\Q$ and the divergence of an axial current, the so-called axial anomaly \cite{Adler:1969gk,Bell:1969ts,Adler:1969er}. 

In the two-component spinor notation  the vector and axial currents in QED/QCD are defined in \p{2.5} and \p{2.6}, respectively.   Using the Feynman rules of Section~\ref{s2}, we evaluate the mixed correlator of two vector and one axial currents, at Born (i.e., free) level,  to be (see also \cite{Schreier:1971um,Erdmenger:1996yc})
\begin{align}\label{6.1}
&\vev{V_\mu(x_1) V_\nu(x_2) A_\la(x_3)}_{\rm Born}  = \frac1{(2\pi^2)^3} \frac{M_{\mu\nu\la}}{x^4_{12} x^2_{13} x^2_{23}}\,, \nt
&M_{\mu\nu\la} = \ep_{\mu' \nu' \rho\la}\, I^{\mu'}_\mu(x_{13}) \, I^{\nu'}_\nu(x_{23}) \, Z^\rho(x_3|x_1,x_2) \,,
\end{align}
where the conformal tensors $I$ and $Z$ have been defined in \p{2.27}. In the interacting theory the vector current  is conserved and hence protected, while the axial one seizes to be conserved due to the axial anomaly. The properly renormalized axial current acquires anomalous dimension starting at two loops \cite{Larin:1993tq}: 
\begin{align}\label{6.2}
\g= - \frac{3 C_F g^4}{2^7 \pi^4} + O(g^6) \,.
\end{align}
 We can predict the following form of the correlator that accounts for the anomalous dimension $\g(g)$ of  $A$ at the point $x_3$ and also for the  beta function:
 \begin{align}\label{6.3}
\vev{V_\mu(x_1) V_\nu(x_2) A_\la(x_3)}_{\rm loop}  = \frac{C(g)}{(2\pi^2)^3} \frac{M_{\mu\nu\la}}{(x^2_{12})^{2-\g/2} (x^2_{13})^{1+\g/2} (x^2_{23})^{1+\g/2}}  +\frac{ \b(g)}{g} \, \Delta_{\mu\nu\la}\,, 
\end{align}
where $C(g)=1+ O(g^2)$. 

The form \p{6.3}  seems to contradict the literature \cite{Schreier:1971um,Crewther:1972kn,Erdmenger:1996yc} where it is  claimed that the only allowed conformal form is \p{6.1}. This assumes that the axial current has canonical dimension. What we have shown in \p{6.3} are two ways of deviating from the form \p{6.1}. The first is possible in a conformal theory with $\b(g)=0$ but with a non-conserved axial current due to the axial anomaly. The second term $\Delta$ is the non-conformal correction due to the non-vanishing beta function.  

{This three-point function satisfies the vector current conservation,}\begin{align}\label{64}
\pa^\mu_1 \, \vev{V_\mu(x_1) V_\nu(x_2) A_\la(x_3)}_{\rm loop} =
\pa^\nu_2 \, \vev{V_\mu(x_1) V_\nu(x_2) A_\la(x_3)}_{\rm loop} = 0\,.
\end{align}
However, due to the anomalous dimension \p{6.2}, the axial current at point 3 is not conserved anymore: 
\begin{align}\label{6.4}
 \vev{V_\mu(x_1) V_\nu(x_2)\,  \pa^\la_{x_3} A_\la(x_3)}_{\rm loop}  = \frac{3 C_F g^4}{2^9 \pi^{10}}\,  \frac{\ep_{\mu\nu\la\rho} x^\la_{13} x^\rho_{23}}{x_{12}^4 x_{13}^4 x_{23}^4} + O(g^6)\,.
\end{align}
What about the conformal symmetry breaking term $\Delta$ on the rhs of \p{6.3}? Without knowing its explicit form, we can argue that it must be conserved. Indeed, the all-order mechanism of non-conservation of the axial current \cite{Adler:1969er} relies on the presence of Adler's fermion triangle subgraph. This has already been accounted for by the first term on the rhs of \p{6.3}, so the same subgraph cannot contribute to the term $\Delta$. 

Further, the Adler-Bardeen theorem \cite{Adler:1969er} tells us that the axial anomaly takes the form of an operator relation between the (properly renormalized) divergence of the axial current and the topological term:
\begin{align}\label{6.5}
\pa^\la A_\la = \frac{g^2}{8\pi^2}\, \Q\,,
\end{align} 
where the coefficient is one-loop exact. Substituting this relation in \p{6.4}, we find exact agreement with our result for the correlator \p{LeviChivita}. 

This simple argument not only confirms the well-know Adler-Bardeen relation \p{6.5} but also gives us an alternative explanation why the correlator \p{LeviChivita} cannot be conformal. Indeed, taking the divergence of a vector of non-canonical dimension in \p{6.4} is not a conformal operation, hence the operator $\Q$ cannot be a conformal primary, as we have shown earlier.

\section*{Acknowledgments}

We are indebted to S. Ferrara, G. Korchemsky, M. Porrati, I. Todorov and A. Zhiboedov for numerous discussions. E.S. is grateful to the MPP-Munich for hospitality during the work on this project.

\appendix

\section{Conventions and conformal properties in position space}
\label{AppConv}

We use the two-component spinor conventions of \cite{Galperin:2001,Galperin:2001uw}. The relations between Lorentz four-vectors and $2\times 2$ matrices are defined by
\begin{align}\label{}
x_{\a\da}=x^\mu(\sigma_\mu)_{\a\da}\,, \qquad \tilde x^{\da\a} = x^\mu(\tilde\sigma_\mu)^{\da\a} = \ep^{\a\b} \ep^{\da\db} x_{\b\db}\,,
\end{align}
with the sigma matrices $\sigma_\mu = (1,\vec\sigma)$ and $\tilde\sigma_\mu = (1,-\vec\sigma)$.
To raise and lower two-component indices we use  the Levi-Civita tensors
\begin{align}\label{}
\ep_{12}=-\ep^{12}=
\ep_{\dot{1}\dot{2}}=-\ep^{\dot{1}\dot{2}}=1\, ,\qquad \ep^{\a\b} \ep_{\b\gamma}=\delta^\a_\gamma  \,,
\end{align}
satisfying the  identities
\begin{align}\label{}
x_{\a\da} \tilde y^{\da\b} + y_{\a\da} \tx^{\da\b} = 2( x \cdot y) \delta_\a^\b\,, \qquad\;
x_{\a\da} \tx^{\da\b} = x^2 \delta_\a^\b\,, \qquad 
x^2 = \frac1{2}  x_{\a\da} \tx^{\da\a}\,.
\end{align}
The space-time derivative is defined as $\pa_{\a\da} = \sigma^\mu_{\a\da} \pa_\mu$ and has the property
\begin{align}\label{A4}
\pa_{\a\da} \tx^{\db\b} = 2 \delta_\a^\b \delta_\da^\db \,.  
\end{align}
Under complex conjugation the Lorentz tensors transform as follows,
\begin{align}
\left[ \ep_{\a\b} \right]^* =\ep_{\da\db} \,,\quad \left[ x_{\a\da} \right]^* = x_{\a\da} \,,\quad 
\left[ (x \tilde y)_{\a}{}^{\da} \right]^* = - (\tilde x y)^{\da}{}_{\a} \,. \label{CC}
\end{align}

The easiest way to check conformal invariance is to make the discrete operation of conformal inversion
\begin{align}\label{}
I[x^\mu] =\frac{x^\mu}{x^2}  \,, \qquad I^2 = \mathbb{I}\,. \label{inv}
\end{align}
In the spinor notation the inversion  acts as follows 
\begin{align}
I[x^{\a\db}_i] = (x^{-1}_i)^{\da\b}\,,\; \,
I[x_{ij}^{\a\db}] = - (x_i^{-1} x_{ij} x_j^{-1})^{\da\b} \,,\; \,
I[\, (x_{ij} \tilde x_{jk})_{\a\b}\,] =  (x_i^{-1} x_{ij} \tilde x_{jk} \tilde x_k^{-1})_{\da\db} \,, \label{CT}
\end{align}
where $x_{ij} \equiv x_i - x_j$. The basic fields in a $D=4$ conformal theory transform with specific conformal weights:
\begin{align}\label{B2}
I[\phi] =   x^2 \phi\,, \quad I[\psi_\a] = x^2 \tx^{\da\a} \psi_\a\,, \quad I[\bar\psi^\da] = -x^2 x_{\a\da} \bar\psi^\da\,, \quad I[F_{\a\b}] =x^2 \tx^{\da\a} \tx^{\db\b} F_{\a\b} \,,
\end{align}
namely $(+1)$ for a scalar $\phi$, $(+3/2)$ for a spinor $\psi$ and $(+2)$ for a field strength. 
These weights are chosen so that the free field equations are covariant. 

The conformal boost generator $\mathbb{K}_\mu$ extends the Poincar\'{e} group to the conformal group. It can be represented as a sequence of an inversion \p{inv}, an infinitesimal space-time translation, and another inversion, i.e. $\mathbb{K}_\mu = I\, \mathbb{P}_\mu\, I$. It acts on an $n$-point correlation function as the following differential operator  
\begin{align}
\mathbb{K}_{\a\da} = i\sum_{i=1}^n \left[ x_i^2 \pa_{i,\a\da} - x_{i,\a\da} x_{i,\b\db} \pa_{i}^{\db\b} -2 \Delta_i x_{i,\a\da} + R_{i,\a\da} \right] \,,  \label{Kboost}
\end{align}
where the Lorentz rotation part $R_{\a\da}$ of the generator acts on the dotted and undotted spinor indices as follows
\begin{align}
&\left[ R_{\a\da}\, \psi \right]_{\beta} =
\left[ R_{\a\da} \right]_{\beta}{}^{\gamma} \psi_{\gamma}= \left[  x_{\beta\da} \delta^{\gamma}_{\a} + \ep_{\beta\alpha} x_{\dot\alpha}^{\gamma}{} \right]\psi_{\gamma} \,,\notag\\
&\left[ R_{\a\da}\, \bar\psi \right]_{\dot\beta} =
\left[ R_{\a\da} \right]_{\dot\beta}{}^{\dot\gamma} \bar\psi_{\dot\gamma}= \left[  x_{\alpha\db} \delta^{\dot\gamma}_{\da} + \ep_{\dot\beta\dot\alpha} x_{\alpha}^{\dot\gamma}{} \right]\bar\psi_{\dot\gamma}\,.
\end{align}



\begin{thebibliography}{99}


\bibitem{Braun:2003rp}
  V.~M.~Braun, G.~P.~Korchemsky and D.~Müller,
  ``The Uses of conformal symmetry in QCD,''
  Prog.\ Part.\ Nucl.\ Phys.\  {\bf 51} (2003) 311
  [hep-ph/0306057].

\bibitem{Poland:2018epd}
  D.~Poland, S.~Rychkov and A.~Vichi,
  ``The Conformal Bootstrap: Theory, Numerical Techniques, and Applications,''
  Rev.\ Mod.\ Phys.\  {\bf 91} (2019) 015002
  [arXiv:1805.04405 [hep-th]].

\bibitem{Larin:1993tq}
  S.~A.~Larin,
  ``The Renormalization of the axial anomaly in dimensional regularization,''
  Phys.\ Lett.\ B {\bf 303} (1993) 113
  [hep-ph/9302240].

\bibitem{Siegel:1979wq}
  W.~Siegel,
  ``Supersymmetric Dimensional Regularization via Dimensional Reduction,''
  Phys.\ Lett.\  {\bf 84B} (1979) 193.

\bibitem{Bern:1995db}
  Z.~Bern and A.~G.~Morgan,
  ``Massive loop amplitudes from unitarity,''
  Nucl.\ Phys.\ B {\bf 467} (1996) 479
  [hep-ph/9511336].
  
\bibitem{Eden:2000mv}
  B.~Eden, C.~Schubert and E.~Sokatchev,
  ``Three loop four point correlator in N=4 SYM,''
  Phys.\ Lett.\ B {\bf 482} (2000) 309
  [hep-th/0003096].
  
\bibitem{Eden:2010zz}
  B.~Eden, G.~P.~Korchemsky and E.~Sokatchev,
  ``From correlation functions to scattering amplitudes,''
  JHEP {\bf 1112} (2011) 002
  [arXiv:1007.3246 [hep-th]].

\bibitem{Eden:2011ku}
  B.~Eden, P.~Heslop, G.~P.~Korchemsky and E.~Sokatchev,
  ``The super-correlator/super-amplitude duality: Part II,''
  Nucl.\ Phys.\ B {\bf 869} (2013) 378
  [arXiv:1103.4353 [hep-th]].

\bibitem{Eden:2012tu}
  B.~Eden, P.~Heslop, G.~P.~Korchemsky and E.~Sokatchev,
  ``Constructing the correlation function of four stress-tensor multiplets and the four-particle amplitude in N=4 SYM,''
  Nucl.\ Phys.\ B {\bf 862} (2012) 450
  [arXiv:1201.5329 [hep-th]].

\bibitem{Chicherin:2020azt}
  D.~Chicherin, J.~M.~Henn, E.~Sokatchev and K.~Yan,
  ``From correlation functions to event shapes in QCD,''
  arXiv:2001.10806 [hep-th].


\bibitem{Grozin:2005yg}
  A.~Grozin,
  ``Lectures on QED and QCD,''
  In *Grozin, Andrey: Lectures on QED and QCD* 1-156
  [hep-ph/0508242].

\bibitem{Schreier:1971um}
  E.~J.~Schreier,
  ``Conformal symmetry and three-point functions,''
  Phys.\ Rev.\ D {\bf 3} (1971) 980.


\bibitem{Erdmenger:1996yc}
  J.~Erdmenger and H.~Osborn,
  ``Conserved currents and the energy momentum tensor in conformally invariant theories for general dimensions,''
  Nucl.\ Phys.\ B {\bf 483} (1997) 431
  [hep-th/9605009].


\bibitem{Drummond:2007au}
  J.~M.~Drummond, J.~Henn, G.~P.~Korchemsky and E.~Sokatchev,
  ``Conformal Ward identities for Wilson loops and a test of the duality with gluon amplitudes,''
  Nucl.\ Phys.\ B {\bf 826} (2010) 337
  [arXiv:0712.1223 [hep-th]].

\bibitem{Adler:1969gk}
  S.~L.~Adler,
  ``Axial vector vertex in spinor electrodynamics,''
  Phys.\ Rev.\  {\bf 177} (1969) 2426.

\bibitem{Bell:1969ts}
  J.~S.~Bell and R.~Jackiw,
  ``A PCAC puzzle: $\pi^0 \to \gamma \gamma$ in the $\sigma$ model,''
  Nuovo Cim.\ A {\bf 60} (1969) 47.

\bibitem{Adler:1969er}
  S.~L.~Adler and W.~A.~Bardeen,
  ``Absence of higher order corrections in the anomalous axial vector divergence equation,''
  Phys.\ Rev.\  {\bf 182} (1969) 1517.
  
\bibitem{Crewther:1972kn}
  R.~J.~Crewther,
  ``Nonperturbative evaluation of the anomalies in low-energy theorems,''
  Phys.\ Rev.\ Lett.\  {\bf 28} (1972) 1421.

\bibitem{Galperin:2001}
A.~Galperin, E.~Ivanov, S.~Kalitsyn, V.~Ogievetsky and E.~Sokatchev,
  ``Unconstrained N=2 Matter, Yang-Mills and Supergravity Theories in Harmonic Superspace,''
  Class.\ Quant.\ Grav.\  {\bf 1} (1984) 469.
\bibitem{Galperin:2001uw}
  A.~S.~Galperin, E.~A.~Ivanov, V.~I.~Ogievetsky and E.~S.~Sokatchev,
  ``Harmonic superspace,''
  Cambridge, UK: Univ. Pr. (2001) 306 p

\end{thebibliography}
\end{document}